\documentstyle[12pt,epsfig]{article}
\begin{document}

\newcommand \be  {\begin{equation}}
\newcommand \bea {\begin{eqnarray} \nonumber }
\newcommand \ee  {\end{equation}}
\newcommand \eea {\end{eqnarray}}

\title{\bf A LANGEVIN APPROACH TO STOCK MARKET FLUCTUATIONS AND CRASHES}

\vskip 2 true cm

\author{Jean-Philippe Bouchaud$^{1,2}$, Rama Cont$^{1,2}$}

\date{{\it\small
$^1$ Service de Physique de l'\'Etat Condens\'e,
 Centre d'\'etudes de Saclay, \\ Orme des Merisiers, 
91191 Gif-sur-Yvette Cedex, France \\ 
%\hbox{  }  \hfill \break
$^2$ Science \& Finance, 109-111 rue Victor-Hugo, 92532 France\\}
\vspace{0.5cm}\today }

\thispagestyle{empty}
\maketitle

\begin{abstract}
We propose a non linear Langevin equation as a model for stock market fluctuations and
crashes. This equation is based on an identification of the different processes
influencing the demand and supply, and their mathematical transcription. We 
emphasize the importance of feedback effects of price variations onto themselves. Risk
aversion, in particular, leads to an `up-down' symmetry breaking term which is
responsible for crashes, where `panic' is self reinforcing. It is
also responsible for the sudden collapse of speculative bubbles. Interestingly,
these crashes appear as rare, `activated' events, and have an exponentially small probability
of occurence. We predict that the `shape' of the falldown of the
price during a crash should be logarithmic. The normal regime, where the stock price
exhibits behavior similar to that of 
 a random walk, however reveals non trivial correlations on different time
scales, in particular on the time scale over which operators perceive a change of
trend. 

\end{abstract}

\begin{center}
 Electronic addresses : \\
bouchaud@amoco.saclay.cea.fr\\
cont@ens.fr\\

\end{center}

\newpage

\section{Introduction}

Stock market fluctuations exhibit several statistical peculiarities which are still
awaiting for a satisfactory interpretation.
More strikingly, many of  
these statistical properties are common to a wide variety of
markets and instruments.
The most prominent features
 are
\cite{reffluct,book,rama,vol,epl,these}:
\begin{enumerate}
\item On short time scales, the variation of stock prices are strongly
non-Gaussian.

\item Market `volatility' (i.e. the variance of the fluctuations) is itself time
dependent, with a slowly decreasing, power-law like, correlation function. 

\item On very long time scales, the log of the price tends to grow linearly with
time, with rare, large drops corresponding to market crashes. 
\end{enumerate}

The first two properties are observed on a certain range of time scales,
ranging from an hour to several weeks, but do not hold for very large time scales (several years) where macroeconomic factors enter into consideration
nor for very short time scales (minutes or so,
the typical duration of a transaction) where the detailed structure of the market has to be taken into account.

These anomalous events have drawn considerable attention recently, both because of
their intrinsic importance, but also because of possible analogies with 
 physical
phenomena such as earthquakes or avalanches. The point is that crashes correspond to
a collective effect, where a large proportion of the actors in a market decide
simultaneously to sell their stocks; it is thus tempting to think of a crash as some
kind of critical point where (as in statistical physics models undergoing a phase
transition) the response to a small external perturbation becomes infinite, because
all the subparts of the system respond cooperatively. Correspondingly, it has been
suggested that  `crash precursors' might exist, and in particular `log-periodic'
oscillations before the crash \cite{logp}. However, no microscopic model has been
proposed which  substantiate such a claim. Actually, there are as yet no convincing
model which `explains' the statistical features described above, although many
proposals have been put forward \cite{models1,models2}. 

A crucial ingredient in model building is the specification  of  the level
(in our case, the time scale) at which one aims to describes the properties
of the system.
There are currently two  major approaches to market dynamics in the economics
and finance literature. One approach is a ``temporary equilibrium"
 approach which assumes that
supply and demand equilibrate quickly enough to be considered at instantaneous
equilibrium at all times \cite{models1}. 
The other one is that of market microstructure theory \cite{ohara}
which examines the implications of market structure, behavorial
assumptions about market participants and
specific trading rules on price behavior at the transaction level.

However, our aim here is to describe market dynamics at time scales 
where, according to empirical observations, some interesting regularities
which are {\it common} between markets with {\it different} microstructures appear \cite{book,these}. At the same time, these time scales
are not long enough to allow the market to reach equilibrium: empirical 
studies show that at intraday time scales there is an imbalance
between supply and demand.
The level of description adopted here is therefore
intermediate between the macroeconomic level
which is that of the market equilibrium models \cite{models1}
and the individual agent level
which is that of the market microstructure theory \cite{ohara}.

The aim of this paper is to propose an alternative description of
the dynamics of speculative markets with
a simple Langevin equation. This equation is built from general
arguments, encapsulating what we believe to be the essential ingredients; in
particular, the feedback of the price fluctuations on the behaviour of the market
participants. We try to motivate as much as possible each term in the equation, and
the value of the corresponding parameters is estimated by comparing with empirical
data. Our basic idea is that although the modelling of each individual participants
(`agents') is impossible in quantitative terms, the collective behavior of
the market and it's impact on the price in particular can be represented
in statistical terms  by a few number of terms
in a (stochastic) dynamical equation. Our approach is in the spirit of many
phenomenological, `Landau-like' approaches to physical phenomena 
\cite{hwakardar}.

\section{A phenomenological Langevin equation}

We denote the price of the stock at time $t$ as $x(t)$. At any given instant of time,
there is a certain number of `buyers' which we call $\phi_+(t)$ (the {\it demand}) and
`sellers', $\phi_-(t)$ (the {\it supply}). The first dynamical equation describes the
effect of an offset between supply and demand, which tends to push the price up (if
$\phi_+ > \phi_-$) or down in the other case. In general, one can write:
\be
\frac{dx}{dt} = {\cal F}(\Delta \phi) \qquad \Delta \phi :=\phi_+ - \phi_-
\ee
where ${\cal F}$ is an increasing function, such that ${\cal F}(0)=0$. In the
following, we will frequently assume that ${\cal F}$ is linear (or else 
that $\Delta \phi$ is small enough to be satisfied with the first term in the Taylor
expansion of ${\cal F}$), and write 
\be
\frac{dx}{dt} = \frac{\Delta \phi}{\lambda}
\ee
where $\lambda$ is a measure of {\it market depth} i.e. the excess demand
required to move the price by one unit.
When $\lambda$ is high, the market can 
`absorb' supply/demand offsets by very small price changes. Now, we try to construct
a dynamical equation for the supply and demand separately. Consider for example the
number of buyers $\phi_+$. Between $t$ and $t+dt$, a certain fraction of those get
their deal and disappear (at least temporarily). This deal is usually ensured by {\it
market makers}, which act as intermediaries between buyers and sellers. The role of 
market markers is to absorb the demand (and supply) even if these do not match
perfectly. Of course, the market makers will absorb buy orders more quickly if they know
that the number of sellers is high, and vice-versa. The effect of market makers
({\sc mm}) can thus be modelled as:
\be
\left.\frac{d\phi_\pm}{dt}\right|_{{\sc mm}} = -\Gamma_\pm(\phi_\mp) \phi_\pm 
\ee
where $\Gamma$ are rates (inverse time scales). We assume that market makers act
symmetrically, i.e, that $\Gamma_+=\Gamma_-$. To lowest order in $\phi$, we write:
\be
\Gamma(\phi) = \gamma + \gamma' \phi + ...
\ee
On liquid markets, the time scale $1/\gamma$  before which a deal is reached is
short; typically a few minutes (see also below for another interpretation of
$1/\gamma$). 

There are several other effects which must be modeled to account for the time
evolution of supply and demand. One is the spontaneous ({\sc sp}) appearance of new buyers
(or sellers), under the influence of new information, individual need for cash, or particular investment strategies. This can be modelled as a white noise term (not necessarily
Gaussian): 

\be
\left.\frac{d\phi_\pm}{dt}\right|_{{\sc sp}} = m_\pm(t) + \eta_\pm(t)
\ee

where $\eta_\pm$ have zero mean, and a short correlation time
$\tau_c$. $m_\pm$ is the average increase of demand (or supply), which might also
depends on time through the time dependent anticipated return $R(t)$ and the
anticipated risk $\Sigma(t)$. It is quite clear that both these quantities are
constantly reestimated by the market participants, with a strong influence of the
recent past. For example, `trend followers' extrapolate a local trend into the future.
On the other hand, `fundamental analysts' estimate what they believe to be the `true'
price of the stock; if the observed price is above this `true' price, the anticipated
trend is reduced, and vice-versa. In mathematical terms, these effects can be
represented as: \be
R(t) =  R_0 + \alpha \int_{-\infty}^t dt' K_R(t-t') \frac{dx}{dt'} - \kappa (x-x_0)
\ee
where $K_R$ is a certain kernel (of integral one) defining how the past average trend is measured by the 
agents, and $\kappa$ is a mean-reversion force, towards the average (over the fundamental analysts) `true
price' $x_0$ \footnote{Note that $x_0$ is actually itself time dependent, although its
evolution in general takes place over rather long time scales (years).}. 

Similarly, the anticipated risk has a short time scale contribution. It is well known
that an increase of volatility is badly felt by the agents, who immediately increase
their estimate of risk. Hence, we write:
\be
\Sigma(t) =  \Sigma_0 + \beta \int_{-\infty}^t dt' K_\Sigma(t-t')
\left[\frac{dx}{dt'}\right]^2
\ee

Correspondingly, expanding $m_\pm(R,\Sigma)$ to lowest order, one has:
\be
m_\pm = m_{0\pm} +  \alpha_{\pm} \int_{-\infty}^t dt' K_R(t-t') \frac{dx}{dt'} 
+ \beta_{\pm} \int_{-\infty}^t dt' K_\Sigma(t-t')
\left[\frac{dx}{dt'}\right]^2 - \kappa_\pm (x-x_0) \label{Eqmpm}
\ee
where the signs of the different coefficients are set by the observation that $m_+$ is
an increasing function of return $R$ and a decreasing function of risk
$\Sigma$, and vice versa for $m_-$. Eq. (\ref{Eqmpm}) contains the leading order
terms which arise if one assumes that the agents try to
reach a tradeoff between risk and return: the demand for an asset decreases
if is recent evolution shows high volatility and increases if it shows an upward
trend.
This is the case for example if the investors follow a
mean-variance optimisation
scheme with adaptive estimates of risk and return \cite{Markovitz}.

Yet another contribution to the change of demand and supply comes from the existence
of option markets, where traders hedge their option positions by buying or selling
the underlying stock. The Black-Scholes rule for hedging relates the number of stock
to be held to the price of the underlying by a non linear formula \cite{BS}. A change
of price thus leads to an increase in the demand or supply which can also be represented
by  $\alpha_\pm$ and $\beta_\pm$ type of terms, reflecting an average of the
so-called `$\Delta$'s' and the `$\gamma$'s' of the different options \cite{BS}. In particular, the
Black-Scholes hedging strategy is a positive feedback strategy of the trend
following type.

We are now in position to write an
equation for the supply/demand offset $\Delta \phi$ by summing all these different
contributions:
\bea \nonumber
\frac{d \Delta \phi}{dt} &=& - \gamma \Delta \phi + m_0 
+ a \int_{-\infty}^t dt' K_R(t-t')
\frac{dx}{dt'} \cr
&-&  b \int_{-\infty}^t dt' K_\Sigma(t-t')
\left[\frac{dx}{dt'}\right]^2 - k (x-x_0) + \eta(t)\label{fund}
\eea
with $a,b,k > 0$. Note in particular that $b>0$ reflects the fact that agents are
risk averse, and that an increase of the local volatility always leads to negative
contribution to $\Delta \phi$. This feature will be crucial in the following. For
definiteness, we will consider $\eta$ to be gaussian and normalize it as: 
 \be
\langle \eta(t) \eta(t') \rangle = 2\lambda^2
D \delta(t-t')
\ee 
where $D$ measure the susceptibility of the market to the random
external shocks, typically the arrival of information. In principle, $D$
should also depend on the recent history, reflecting the fact that an 
increase in volatility induces a stronger reactivity of the market to external
news. In the same spirit as above, one could thus write:
\be
D = D_0 + D_1 \int_{-\infty}^t dt' K_D(t-t')
\left[\frac{dx}{dt'}\right]^2\label{d1}
\ee
For simplicity, we will neglect the influence of $D_1$ in the following
sections, but comment on its effect in the concluding section.

Finally, let us note that Eq. (\ref{fund}) can be extended to allow for 
agents with different reaction times. For example, the term 
$a \int_{-\infty}^t dt' K_R(t-t') \frac{dx}{dt'}$ can be generalized as :
\be
\sum_i a_i \int_{-\infty}^t dt' K_R^i(t-t') \frac{dx}{dt'}
\ee
where the $K_R^i$ have different ranges. As we shall discuss below, the 
empirical data suggests that there is a population of very fast traders 
(probably market makers themselves) which in a 
contrarian way ($a_i < 0$).

\section{Analysis of the linear theory. Liquid vs. Illiquid markets}

Let us consider the linear case `risk neutral' case where $b=0$. We will assume
for simplicity that $K_R(t) = \Gamma \exp -(\Gamma t)$, and first consider the local
limit where $\Gamma$ is much larger that $\gamma$ (short memory time). In this case, the equation for
$x$  becomes that of an harmonic oscillator \footnote{In an oral seminar given 
in Jussieu in June 1997, Doyle Farmer also presented a second-order equation for the price. We are not aware of the existence of a written version, and do not
know to what extend his analysis is similar to ours.}:
\be
\frac{d^2 x}{dt^2} + \left(\gamma-\frac{a}{\lambda}\right) \frac{dx}{dt} + \frac{k}{\lambda} 
(x - \tilde x_0) = \frac{1}{\lambda} \eta(t) 
\ee
where $m_0$ has been absorbed into a redefinition of $\tilde x_0 := x_0 + m_0/k$. For
liquid markets, where $\lambda$ and $\gamma$ are large enough, the `friction' term $\tilde \gamma := \gamma-a/\lambda$ is positive. In this
case the market is stable, and the price oscillates around an equilibrium value
$\tilde x_0$, which is higher than the average fundamental price if the spontaneous
demand is larger than the spontaneous supply (i.e. $m_0$ is positive), as expected when
the overall economy grows. One can also compute the time correlation function of the
price fluctuations. The important parameter is: 
\be
\epsilon := \frac{k}{\lambda \tilde \gamma^2}
\ee
For liquid markets, $\epsilon \ll 1$. The correlation is found to be the sum of two 
exponentials, with correlation times $\tau_1,\tau_2$:
\be
\tau_1 \simeq \frac{1}{\tilde \gamma} \qquad \tau_2 \simeq \frac{\tau_1}{\epsilon}
\ee
and amplitudes ${\cal A}_{1,2}$ such that ${\cal A}_2 \simeq \epsilon^2 {\cal A}_1$. Thus, on a time
scale $\tau_1$, the correlation function falls to a very small value $\sim
\epsilon^2$. This allows one to identify $\tau_1$ with the correlation time observed
on liquid markets, which is of the order of several minutes \cite{book}, thereby fixing the order
of magnitude of $\tau_1^{-1}=\tilde \gamma \simeq \gamma$. Thus, on time scales such that $\tau_1 \ll t \ll \tau_2$,
the stock price behaves as a simple biased random walk with volatility $\sigma^2 = 2 D
\tau_1^2$, before feeling the confining effect of the `fundamental' price. Since the
fundamental price is surely not known to better than -- say -- $10 \%$, and that the
typical variation of the price of a stock is also around $10 \%$ per year, it is
reasonnable to assume that the time scale $\tau_2$ beyond which `fundamental' effects
(as represented by the harmonic term) play a role is of the order of a year
\footnote{Beyond the year time scale, however, the
evolution of $x_0$ itself cannot be neglected.}.  This leads to $\epsilon \simeq 10^{-4}$. Hence, for liquid markets, the role of the confining term can probably be
neglected, at least on short time scales. 

The situation is rather different for illiquid markets, or when trend following
effects are large, since $\tilde \gamma$ can be negative. In this case, the market is
unstable, with an exponential rise or decay of the stock value, corresponding to a
speculative bubble.  However, in this case, $dx/dt$ grows with time and it soon
becomes untenable to neglect the higher order terms, in particular the risk aversion
term proportional to $b$. We will comment on this case below. Let us however start by
analyzing the role of $b$ for liquid markets for which, as explained above, it is
reasonnable to set $k=0$. 

\section{Risk aversion induced crashes as activated events}

Setting $u=dx/dt$ and still focusing on the limit where the memory time $\Gamma^{-1}$ is very
small, one finds the following non linear Langevin equation:
\be
\frac{du}{dt} = \frac{m_0}{\lambda} -\tilde \gamma u - \frac{b}{\lambda} u^2 +
\frac{1}{\lambda} \eta(t) \equiv -\frac{\partial V}{\partial u} 
+ \frac{1}{\lambda} \eta(t) \label{Langevin1}
\ee
This equation represents the evolution of the position $u$ of a viscous fictitious particle
in a potential $V(u)$ represented in Fig. 1.

\begin{figure}
\centerline{\hbox{\epsfig{figure=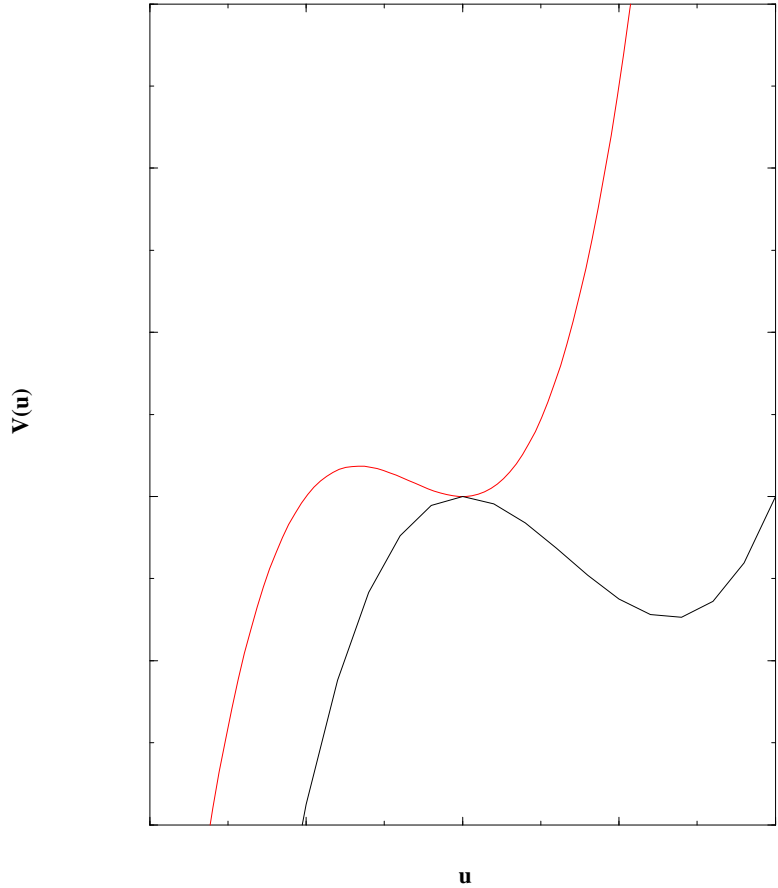,width=8cm}}}
\vskip 0.8cm
\caption{Shape of the effective potential $V(u)$. For liquid markets, $\tilde
\gamma >0$ and $V(u)$ is minimum for $u \propto m_0$. For illiquid markets, or when the `trend following' effect is large, the minimum moves to 
to positive value unrelated to $m_0$. The interesting point is the presence of a potential barrier, separating a normal random walk like regime from a crash regime.} \label{fig1}
\end{figure} 

In order to keep the mathematical form simple, we
set the average trend $m_0/\lambda$ to zero (no net average offset between spontaneous demand and
spontaneous supply); this does not qualitatively change the following picture, unless
$m_0$ is negative and large. The potential $V(u)$ can then be written as: \be V(u) =
\frac{\tilde \gamma}{2} u^2 + \frac{b}{3\lambda} u^3 \ee
which has a local minimum for $u=0$, and a local maximum for $u^* = - \lambda \tilde
\gamma/b$, beyond which the potential plumets to $-\infty$. The `barrier height' $V^*$
separating the stable region around $u=0$ from the unstable region is given by:
\be
V^*=V(u^*)-V(0)= \frac{\tilde \gamma u^{*2}}{6}
\ee
The nature of the motion of $u$ in such a potential is the following: starting at 
$u=0$, the particle has a random harmonic-like motion in the vicinity of $u=0$ until
an `activated' event (i.e. driven by the noise term) brings the particle near $u^*$. Once this barrier is crossed,
the fictitious particle reaches $-\infty$ in finite time. In financial terms, the
regime where $u$ oscillates around $u=0$ and where $b$ can be neglected, is the
`normal' random walk regime discussed in the previous paragraph. (Note that the random walk is biased when $m_0 \neq 0$). This normal regime
can however be interrupted by `crashes', where the time derivative of the price
becomes very large and negative, due to the risk aversion term $b$ which enhances the drop in the price. The point is that these two regimes can be clearly separated since the average time $t^*$ needed for such crashes
to occur can be exponentially long, since it is given by the classical
Arrhenius-Kramers formula \cite{Wax,Hanggi}:
\be
t^* = 2 \pi \tau_1 \exp\left(\frac{V^*}{D}\right) = \frac{2 \pi}{\gamma} 
\exp\left(\frac{u^{*2}\tau_1}{3 \sigma^2}\right)\label{Arrh}
\ee
where $D$ is the variance of the noise $\eta$ and $\tau_1=1/\tilde \gamma$. Taking $t^*=10$ years, $\sigma=1 \%$ per day, and
$\tau_1=10$ minutes, one finds that  the characteristic value $u^*$ beyond which the
market `panics' and where a crash situation appears is of the order of $-1 \%$ in ten
minutes, which not unreasonnable. The ratio appearing in the exponential can also be
written as the square of $u^* \tau_1/\sigma \sqrt{\tau_1}$; it thus compares the value
of what is considered to be an anomalous drop on the correlation time ($u^*\tau_1$) to the
`normal' change over this time scale ($\sigma \sqrt{\tau_1}$). 

Note that in this line of thought, a crash occurs because of an improbable succession
of unfavorable events, and not due to a single large event in particular. Furthermore, there are
no `precursors' - characteristic patterns observed before the crash:
 before $u$ has reached $u^*$, it is impossible to decide whether it
will do so or whether it will quietly come back in the `normal' region $u \simeq 0$. Note finally that an increase in the liquidity  factor 
$\gamma$ reduces the probability of crashes. This is related to the stabilizing role of market makers, which appears very clearly.

An interesting prediction concerns the behaviour of the price once one enters the
crash regime i.e. once $u$ becomes larger (in absolute value) than $u^*$.
Neglecting the noise term, one finds that the stock price is given by:
\be
x(t) = x^* + \frac{\lambda}{b} \ln \left[\exp(\tilde \gamma
(t_f-t))-1\right]\label{log} 
\ee
which diverges logarithmically towards $-\infty$ when $t$ reaches a final time $t_f$.
Of course, in practice, this divergence is not real since when the price becomes too
low, other mechanisms, which we have not taken into account in the model, come into
play (for example the action of a regulating authority).
 One thus expects that some external
mechanism interrupts the crash, which in the Langevin language, correspond to a
`reinjection' of the particle around $u=0$. Formula (\ref{log}) is compared in Fig. 2
to the observed price of the S\&P index during the 1987 October crash, where we have fixed
$t_f$ to be the time when the price reaches its minimum. This leads to $\tilde \gamma
= 4.5 \ 10^{-3}$ (in minutes$^{-1}$) and $\lambda/b = 12.9$ (S\&P points), from which
we estimate $u^*=3.5$ S\&P points per hour (more than $1\%$ per hour). The last
figure is not unreasonable; however, the order of magnitude found for $\tilde \gamma$
is much smaller than expected, on the basis that $\tau_1$ is ten minutes or so. We
shall come back to this point below, in section (\ref{memory})

\begin{figure}
\centerline{\hbox{\epsfig{figure=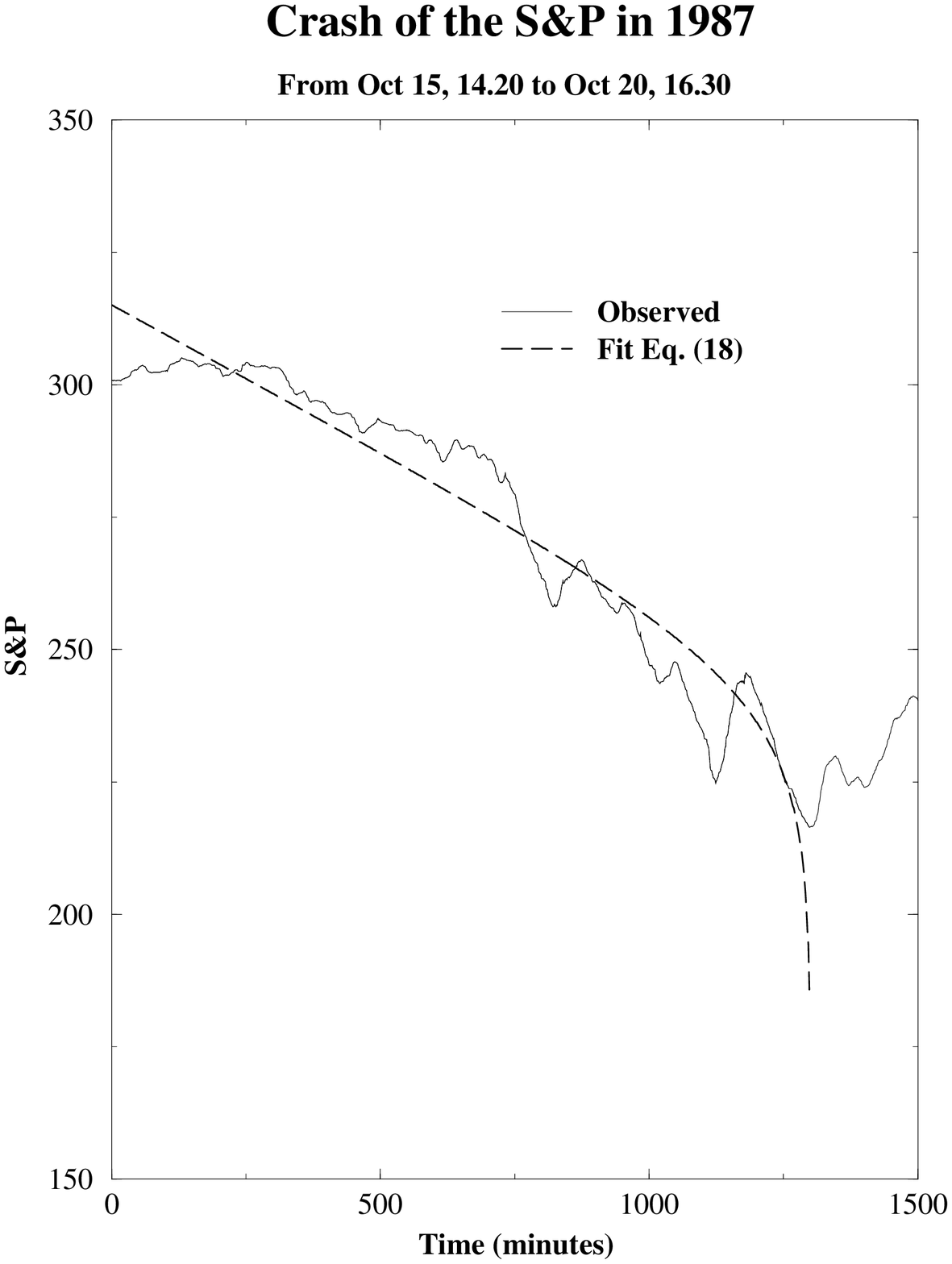,width=8cm}}}
\vskip 0.8cm
\caption{Evolution of the New-York S\&P index during the 1987 October crash. 
The dotted line is a fit with the noiseless formula (\protect\ref{log}), where $t_f$ is taken to be the time when the index reached its minimum. At this point, our model certainly breaks down, since other effects, not taken into account in the
present approach, come into play. } \label{fig2}
\end{figure}

\section{Illiquid markets: speculative bubbles and collapse}

Suppose now that the trend following tendency is strong so that $\tilde \gamma < 0$. 
The potential $V(u)$ has a minimum for $u=u^*$ which is now {\it positive}, and
a maximum for $u=0$. The price increment $u$ oscillates around a
positive value unrelated to $m_0$, which means that there is a non zero 
trend not based on true growth but entirely induced by the fact that a price increase motivates more people to buy --
this is called a speculative bubble. After a time $t$, the price has risen on average
by an amount $u^* t$. If this increase is too large, it becomes illegitimate to
neglect the role of the `fundamental' price $x_0$. The full potential in which $u$
evolves is actually given by $V(u) + k (x-x_0)u/\lambda$. It is easy to see that 
this potential has a local minimum (which leads to the above sustained growth) only when:
\be
\tilde \gamma^2 > 4\frac{kb}{\lambda^2} (x-x_0)
\ee
but that this minimum disappears for larger values of $x-x_0$. Assuming that
$x(t=0)=x_0$, we thus find a time $t_b$ where the bubble has to collapse, since the
(metastable) equilibrium around $u^*$ is no longer present. The lifetime of the
bubble is given by:
\be
t_b \simeq \frac{a -\lambda \gamma}{k}
\ee
As could be expected, the stronger the  damping term pulling the price
back 
towards the fundamental value $x_0$, the
shorter will be the duration of speculative bubbles.

\section{Memory effects}
\label{memory}
Up to now, we have assumed that the impact of a change of price on the behaviour of
the market participants was instantaneous, i.e. that the kernels $K_R$ and
$K_\Sigma$ used to estimate the average return and risk have a typical memory time
shorter than any other time scale in the problem, in particular $\gamma^{-1}$. Since
we have argued that $\gamma^{-1}$ is of the order of a few minutes on liquid markets,
this assumption is not very realistic: it is more reasonable to assume that 
agents judge the evolution of risk and return on longer time scales, at least several
days. We are thus actually in the opposite limit where $\Gamma \ll \gamma$. 
Fortunately the case
of an exponential memory kernel  still leads to a tractable model. It
is easy to show that the dynamical equation now reads:
\be
\frac{d^2u}{dt^2} =
-(\gamma+\Gamma) \frac{du}{dt} - \Gamma\left[\frac{m}{\lambda} -\tilde \gamma u -
\frac{b}{\lambda} u^2 \right] + \frac{1}{\lambda} \left[\Gamma \eta(t)+
\frac{d\eta}{dt}\right] \label{Langevin2}
\ee
which indeed leads back to Eq. (\ref{Langevin1}) in the limit $\Gamma \to \infty$.
Eq. (\ref{Langevin2}) governs the evolution of a massive particle in the very same
potential as the one above (Fig 1). One can show that in this case, for $b=m_0=0$, and in the
limit where $\Gamma \ll \gamma$, the correlation function of the increment $u$ is given
by: \be
\langle u(t) u(t') \rangle = \sigma^2 \left[(\tilde \gamma^2 + \tilde \gamma
\tilde \Gamma) \exp (- \gamma|t-t'|)+ (\tilde \Gamma^2 + \tilde \gamma \tilde \Gamma)
\exp - (\Gamma|t-t'|) \right] 
\ee
(with $\tilde \Gamma = {2\Gamma a}/{\lambda \gamma}$)
which thus decays rapidly (on a time $\tilde \gamma^{-1}$) to a (small) value which can be
negative if $a<0$, before slowly going to zero on time scales $\sim \Gamma^{-1}$.
Interestingly, the empirical correlation function of short time increments indeed
shows a {\it negative} minimum on time scales of the order of 15 minutes \cite{book,these}.
The relative amplitude of this minimum (as compared to $\langle u(t)^2 \rangle$) is of
the order of a few $\%$. This could thus be interpreted as the effect of fast
`contrarian' traders superposed to the regulatory action of the market maker
(contributing to a negative $a$).

For $b>0$, one still has a sharp distinction between a `normal' regime, where the 
stock price performs a random walk with volatility $\sigma$ (except that, as
just discussed, the increment correlation function has a small tail decaying on time
scales $\Gamma^{-1}$), and a `crash' regime, when the `particle' manages to reach the
top of the potential barrier. The theory of activated processes can be extended to
massive particles. In the limit $\Gamma \ll \tilde \gamma$, the average time between
crashes $t^*$ is given by a formula very close to the one above \cite{Wax,Hanggi}:
\be
t^* = \frac{2 \pi}{\sqrt{\tilde \gamma \Gamma}} \exp\left(\frac{u^{*2}\tau_1}{3
\sigma^2}\right) 
\ee
i.e., only the prefactor of the exponential is changed. Note that, as could be
expected intuitively, the fact that there is a delay in the reaction of traders
tends to stabilize the market, since the crash time is multiplied by a factor 
$\sqrt{\tilde \gamma/\Gamma}$. 

Finally, the dynamics of the price when the crash has started is also affected
by the presence of a memory. The truly asymptotic behaviour of Eq. (\ref{Langevin2}) 
(for zero noise) is given by:
\be
u(t) \simeq -\frac{6 \lambda}{\Gamma b} (t_f-t)^{-2}
\ee
which leads to a $(t_f-t)^{-1}$ divergence of the price itself. However, as noticed
above, this divergence is certainly interrupted by effects which our model cannot 
describe. In order to compare with empirical data, in particular that of the crash of
1987, one can notice that the time scale over which the crash took place (days) is
much larger than $\tilde \gamma^{-1}$. It is thus reasonnable to neglect the second 
derivative term as compared to the first. In the limit where $\Gamma \ll  \gamma$, we are thus led to:
\be
\frac{du}{dt} =  -\Gamma u - \frac{b \Gamma}{\lambda \gamma} u^2 
\ee
the solution of which being of the same form as the one without memory, except for
the coefficients:
\be
x(t) = x^* + \frac{\lambda \gamma}{b \Gamma} \ln \left[\exp(\Gamma
(t_f-t))-1\right]\label{log2} 
\ee
The same fit as in Fig. 2 is thus adequate. However, interestingly, one finds that 
it is now $\Gamma$, rather than $\tilde \gamma$, which appears in the exponential. In
other words, the time scale during which the crash develops is much longer; from the
fit we find (see Fig 2): $\Gamma^{-1} \simeq 220$ minutes (half a day). The estimate of
$u^*$, as  $\lambda \gamma/b$, is however unaffected.

\section{Concluding remarks}

We hope to have convinced the reader that the above Langevin equation, which is based
on an identification of the different processes influencing supply and
 demand, and
their mathematical transcription, captures many of the features seen on markets. We
have in particular emphasized the role of feedback, in particular through risk
aversion, which leads to an `up-down' symmetry breaking non linear term $(dx/dt)^2$.
This term is responsible for the appearance of crashes, where `panic' is self
reinforcing; it is also responsible for the sudden collapse of speculative bubbles.
Interestingly, however, these crashes are rare events, which have an exponentially
small probability of occurence (see Eq. (\ref{Arrh})). We predict that the `shape' of
the falldown of the price during a crash should be logarithmic (see Eq.
(\ref{log2})), which is compatible with empirical data (Fig. 2). The `normal' regime,
where the stock price behaves as a random walk, reveals non trivial correlations
on the time scale over which operators perceive a change of trend. In particular, a small negative dip related to the existence of contrarian 
traders can appear.  In this
respect, it is important to stress that within these models lead, in principle, to simple winning strategies. It is however easy to convince oneself that if the level of
correlations is small (for example, as seen above, of order of a few percent after
tens of minutes), the transaction costs are such that arbitrage cannot be implemented
in practice \cite{book}. Therefore, we believe that non trivial correlations can be
observed on financial data, and do actually arise naturally when feedback effects are
included.

Before closing, we would like to discuss briefly several other points. The first one
concerns the fact that we have considered $x$ to be the price, rather than the log of
the price. Of course, on short time scales, this does not matter, and actually a
description in terms of the price itself is often preferable on short time scales
\cite{book}. On longer time scales, however, the log of the price should be prefered since 
it describes the evolution of prices in {\it relative}
rather than absolute terms.
However, on these long time scales, one should also take into account the evolution of
the model's parameters (such as the fundamental price $x_0$, or the average trend $m_0$),
which is related to true economics, and thus not amenable to such a simple statistical
treatment as we have argued for psychology. Second, we have identified a `normal'
regime, where $u$ oscillates around zero, and a crash regime for $|u| > u^*$. In the
model presented above, the `normal' fluctuations are gaussian \footnote{Note that in
our model, `normal' fluctuations and crashes describe two very different regimes of
the same dynamical equation. In this sense, we agree with the idea that market
crashes are indeed `outliers' from a statistical  point of view \cite{sornette}} if $\eta$ is
gaussian and if the relation between price changes and supply/demand unbalance is
linear. In order to account for the large kurtosis observed on markets during `normal'
periods (i.e. excluding crashes), one necessarily has to take into account either the
non linearity of the price change and/or the non-normal nature of the `noise', in particular the role of the feedback term $D_1$ introduced in Eq. (\ref{d1}),
which can indeed be shown to lead to `fat tails' \cite{ustocome}.
Although the quantitative formulae given above are affected by such effects, the
qualitative picture will remain.

Finally, the above model, where crashes appear as
activated events, suggests a tentative interpretation for `log-periodic'
oscillations seen before crashes \cite{logp}. Imagine that each time $u$
reaches -- by accident --  an anomalously negative value (but above $u^*$), the
market becomes more `nervous'. This means that its susceptibility to external
disturbances like news will increase. In our model, this can be described by an
increase of the parameter $D \propto \sigma^2$, through the term $D_1$ in Eq.
(\ref{d1}). If  $D$ increases by 
a certain value
$\delta D$ at every accident and since $D$ appears in an exponential, this implies
that the average time $\Delta t$ before the next `accident' is decreased by a certain factor
which, to linear order in $\delta D$, is constant:
\be
\Delta t_{n+1} = \Delta t_{n} S^{-\delta D/D} \qquad S=\exp\left(\frac{u^{*2}\tau_1}{3
\sigma^2}\right). 
\ee
This leads to a roughly log-periodic behaviour, which indeed predicts that the time difference
between two events is a geometric series. However, our scenario is {\it not} related to a critical point: the crash appears when $u$ exceeds $u^*$, and not when $\Delta t \to 0$, i.e., when crash events
accumulate. In this respect, it should be noted that according to the critical log
periodic theory, there should have been another crash near the end of November 1997, and
then again roughly 10 days later, which did not occur \cite{lalouxpotters}.   

\vskip 1cm

{\sc Acknowlegments.} We would like to thank J.P. Aguilar, S. Galluccio,
 L. Laloux and especially M.
Potters for many discussions on the problem of stock market fluctuations and crashes.

\end{document}